# Fundamental Limitations in Biomarker Based Early Disease Diagnosis

*Tuhin Chakrabortty[*], Chandra R. Murthy[$], Manoj M Varma[*#]*
*Center for Nano Science and Engineering, Indian Institute of Science, Bangalore, India
$Department of Electrical Communication Engineering, Indian Institute of Science, Bangalore, India
# Robert Bosch Centre for Cyber-Physical Systems, Indian Institute of Science, Bangalore, India

**Abstract:**
Clinical biosensors with low detection limit hold significant promise in the early diagnosis of debilitating diseases. Recent progress in sensor development has led to the demonstration of detection capable of detecting target molecules even down to single molecule level. One crucial performance parameter which is not adequately discussed is the issue of measurement fidelity in such sensors. We define measurement fidelity based on the false positive rate of the system as we expect systems with higher sensitivity to concomitantly respond more to interfering molecules thus increasing the false positive rates. We present a model which allows us to estimate the limit of detection of a biosensor system constrained by a specified false-positive rate. Two major results emerging from our model is that a) there is a lower bound to the detection limit for a target molecule determined by the variation in the concentration of background molecules interfering with the molecular recognition process and b) systems which use a secondary label, such as a fluorophore, can achieve lower detection limits for a given false positive rate. We also present data collected from literature to support our model. The insights from our model will be useful in the systematic design of future clinical biosensors to achieve relevant detection limits with assured fidelity.

**Introduction:**

Early diagnosis of several diseases significantly improves the subsequent prognosis an example of which is the exponential increase in the 5-year survival probabilities in certain types of cancers [1]. Early diagnosis relies on successfully detecting the presence of biomarker molecules which may often be 9-10 orders of magnitude lower in concentration compared to other molecules in complex multi-component samples such as serum [2]. Over the last several decades, various groups have demonstrated bio-molecular sensing techniques which have advanced to the point of detecting the presence of a few molecules in a given sample [3]. These techniques have relied on sensor miniaturization coupled with sensitive detection of small signal shifts due to molecular binding. A notable example of this approach is Digital ELISA (d-ELISA) which can detect the presence of a few specific molecules in serum, a sample presenting a complex background [4]. While standard ELISA employs macroscale entities such as 96-well plates, d-ELISA miniaturizes the bio-recognition reaction within a single micron (2.7 $\mu$m) sized bead along with an optical fiber-based readout to detect single-bead resolved bio-recognition reaction with high sensitivity. Thus, the standard ELISA system gets digitized to a format where the number of binding events on every bead can be counted as a discrete signal. Similarly, other groups have also attempted to use novel read-out techniques to improve the detection limit of the biosensors. For example, Guider et al. [5] used micro-ring resonators for biomolecular detection. Similarly, Puczkarski et al. used graphene nanoelectrodes as the read-out technique for biomolecular sensing [6]. While clever techniques like these have addressed the topic of achieving lower limits of detection for biomarkers, relatively far fewer studies have investigated the issue of measurement fidelity in biosensors. Measurement fidelity refers to the accuracy of measurement, i.e. a set of measurements with perfect fidelity would only consist of true positives or true negatives. The issue of fidelity is of paramount importance because a) it is independent of the detection limit, i.e. the ability to measure low-levels of signals doesn't imply high fidelity and vice-versa and b) practical utility of a sensor is largely limited by its fidelity, for instance, a moderately sensitive (within acceptable limits) biosensor with a low false-positive rate is likely to be preferable than a single molecule sensor with an unacceptably large false-positive rate. From the clinical point of view high false-positive rates would result in un-justified medical costs for further diagnostics, psychological stress and erosion of user-confidence in the test results, all of which are undesirable. Therefore, systematic studies of measurement fidelity in biosensing systems is of crucial importance.

There is a significant amount of theoretical work in the literature on the precision of chemical sensing (equivalent to our notion of measurement fidelity) in the case of living sensory systems such as single cells in the presence of spurious ligand molecules [7, 8, 9, 10, 11]. However, clinical biosensors are not limited by the assumptions used to describe the performance of living systems such as biological cells because experimental factors such as the read-out technique, concentrations of receptors, and serum can be manipulated in the case of clinical biosensors presenting scenarios which cannot be properly described by equivalent models for biological cells. Because of the increased degrees of freedom in the case of clinical biosensors relative to sensing by a biological cell, it is crucial to develop independent theoretical models to understand the effects of controllable experimental factors for desirable system performance such as low detection limits and/or high measurement fidelity. From the point of view of describing measurement fidelity in clinical biosensors, the most important prior work is the stochastic binding model with generalized amplification stages presented by Hassibi et al [12]. Although their paper does not explicitly refer to the notion of measurement fidelity, one of the results they obtain, namely that the detection limit reaches a lower bound with increasing background is reminiscent of the one of the two main results obtained in this article although in a different context as discussed later. The analysis presented by Hassibi et. al is very generic, which allows their method to be applied universally, but unfortunately the generalized approach also makes it more difficult to relate experimental processes and associated parameters to the generalized model. There is a lack of appropriate mathematical models describing measurement fidelity of clinical biosensors which motivates us to present such a model. In this article, we derive analytical expressions for the limit of detection of clinical biosensors with measurement fidelity specified as a prescribed, acceptable rate of false positives. The first major result revealed by our analysis is the presence of a lower bound for the detection limit (for a specified fidelity level) depending on uncertainty in parameter values such as the background concentration. Further, our analysis shows that the choice of read-out technique has a strong influence on the detection limit of a biosensor. This is because read-out techniques strongly affect the signal background which in-turn affects the fidelity. We classify the read-out techniques into labelled or un-labeled depending on the use of labels such as fluorophores, chromophores, magnetic particles etc. in the detection process as illustrated schematically in Fig. 1. The second major result emerging from our model is that label-based detection techniques would have better signal fidelity than un-labeled methods for the same requirements. We demonstrate the validity of our claims by comparing our model predictions with data collected from the literature.

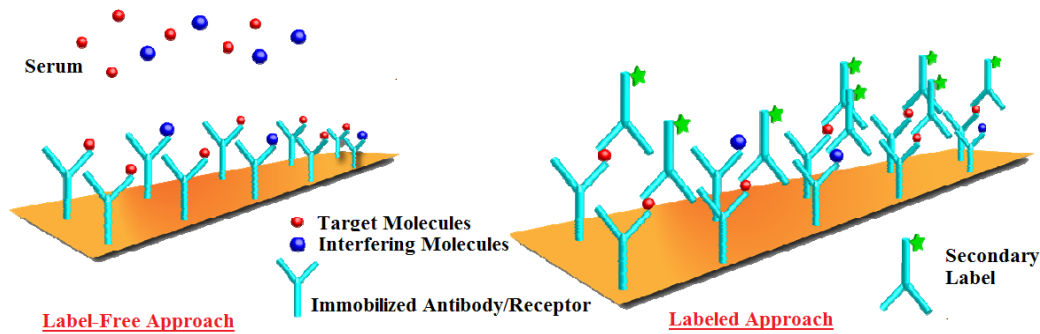

**Figure 1** Classification of read-out technique: Fig. (a) shows the un-labeled approach where the binding of the target molecule to the receptor directly produces the signal. Fig. (b) shows the labeled approach where a secondary label is used to produce a signal (e.g. fluorescence) from the target molecules bound to the receptors. As the signal only arises from the bound molecules, signal background arising from unbound receptors, as discussed in the main text, is essentially zero.

**Mathematical model:**

Detection of disease marker molecules using clinical biosensors typically involve a surface-based chemical recognition system [13]. Here, the surface of a solid substrate, such as a glass slide, is coated with molecules which serve as receptors for the target molecules, i.e. the receptors on the surface bind target molecules in the clinical sample. In the extreme limit, an ideal receptor should possess infinite specificity, i.e. they should only bind the target molecules (which we refer to as specific molecules in the rest of the article) and no other molecular species. Therefore, in theory, when a sample containing a large number of different molecular species is poured onto the sensor surface, only the specific molecules, if present in the sample, will get attached to the surface and as a result, some measurable properties of the surface like refractive index, electrical conductivity would change.

Clinical biosensors most often use proteins called antibodies as receptors and these receptors are generally polyreactive, i.e. they have a binding affinity for several unrelated antigens [14] Therefore, one also needs to consider the effect of binding of non-specific spurious molecules while estimating the concentration of biomarkers. To account for this in a typical experimental set up, a sensor is first calibrated using a negative control, i.e. a sample not containing the specific target biomarker. The output of this measurement serves as a baseline for the detection. The same sensor or an identical one is then used for probing the sample suspected to be containing the specific biomarker. The difference in the output signals from the two measurements indicates the presence of the biomarker in the sample. The sample is typically serum, a form of purified blood containing a large number of molecular species such as proteins, signaling molecules, metabolites etc. in varying concentrations [2]. It is conceivable that some of these may also bind to the receptors inducing spurious signals during the measurement process.

Let $S$ and $S_0$ be the output signals from experiments in presence and absence of the specific target biomarker respectively, then $S$ and $S_0$ can be written as

$$S = \rho_b b + \rho_u(C_r - b) + \xi$$
$$S_0 = \rho_b b_0 + \rho_u(C_r - b_0) + \xi \qquad (1)$$

where $C_r$ is the total receptor concentration and $\rho_b$ and $\rho_u$ are the signal strengths due to the bound and unbound receptors, respectively, and $\xi$ is the measurement noise which is a property of the instrument and is independent of the molecular reaction parameters, for instance, the fluctuation in the intensity of the light source for a refractive index measurement or voltage fluctuations in the case of impedance measurements. We assume $\xi$ to be a Gaussian distributed variable with $\langle \xi_i \rangle = 0$ and $\langle \xi_i \xi_j \rangle = \sigma_\xi^2 \delta_{ij}$. Variables $b$ and $b_0$ are respectively, the average concentrations of the bound receptors at equilibrium in presence and absence of the specific biomarkers and can be calculated from the chemical kinetic equations. In a typical experimental setting, one would like the concentration of the receptors to be significantly higher compared to the other species to avoid signal saturation. Signal saturation occurs when all receptors are bound by either the specific target biomarker, denoted by subscript 's', or any non-specific molecule, denoted by subscript 'n' in this article. If $C_s$ and $C_n$ are the concentration of specific and non-specific target molecules respectively, then for $C_r \gg C_s + C_n$, the equations for the fraction of bound receptors due to specific and non-specific target molecules can be written as

$$\frac{db_s}{dt} = k_{on-s}(C_s - b_s)(C_r - b_s - b_n) - k_{off-s}b_s$$
$$\frac{db_n}{dt} = k_{on-n}(C_n - b_n)(C_r - b_s - b_n) - k_{off-n}b_n \qquad (2)$$

Where $k_{on-i}$ and $k_{off-i}$ are the binding and unbinding rates.

At equilibrium, the expressions for $b_0$ and $b$ become

$$b_0 = b_n = \frac{C_r C_n}{C_r + K_{Dn}}$$
$$b = b_n + b_s = \frac{C_r C_s}{C_r + K_{Ds}} + \frac{C_r C_n}{C_r + K_{Dn}} \qquad (3)$$

$K_{Di}$ is the dissociation constant of the receptor with the respective biomolecule which is defined as $K_{Di} = \frac{k_{off-i}}{k_{on-i}}$, where $k_{off}$ and $k_{on}$ are the unbinding and the binding rate of the biomolecules. For $C_r \gg K_{Dn}$, equation (3) becomes

$$b_0 = C_s$$
$$b = C_s + C_n \qquad (4)$$

Other than the measurement noise ($\xi$), there are several experimental factors that can also act as the source of noise for the clinical biosensors. For example, concentration of the non-specific biomolecules in the serum may vary significantly from person to person. Therefore, there will be observable fluctuations in the baseline calibration. Furthermore, all the receptors coated on the surface may not bind to the surface properly. As a result, the concentrations of these receptor molecules are also known only to a finite precision. Additionally, fluctuations in the environmental conditions, such as temperature and pH, will change the binding affinities and thereby the dissociation constants of the reactions. Therefore, fluctuations in these parameters need to be taken into consideration when calculating precision of measurement. In our analysis the effect of fluctuations in the dissociation constant ($K_D's$) is neglected as experiments can be carried out with reasonably stable temperature and pH conditions. However, fluctuations in receptor concentration ($C_r$) and concentration of non-specific molecules ($C_n$) will be significant due to distribution of orientation of the antibody receptors, variations in the background molecule concentration and so on [15]. We take $C_r$ and $C_n$ to be Gaussian distributions as follows

$$C_r = \overline{C_r}(1 + \mathcal{N}(0, \delta_r^2))$$
$$C_n = \overline{C_n}(1 + \mathcal{N}(0, \delta_n^2))$$

where $r$ and $n$ represent receptor molecules and non-specific molecules respectively. $\bar{C}$ is the average concentration of the respective molecules and $\delta$ is the relative fluctuation from the average defined as

$$\delta = \frac{\sigma}{\bar{C}}$$

here $\sigma^2$ is the variance of the concentration.

The difference in the output signals $S$ and $S_0$ can now be calculated as

$$S - S_0 = \Delta\rho C_s + \sqrt{2}\big(\Delta\rho\mathcal{N}(0, \sigma_n^2) + \rho_u\mathcal{N}(0, \sigma_r^2)\big) + \sqrt{2}\xi$$

Where $\Delta\rho = \rho_b - \rho_u$

The problem of detection now becomes a hypothesis testing problem where the null hypothesis ($H_0$) and the alternative hypothesis ($H_1$) can be defined as
$$H_0: S - S_0 = 0$$
$$H_1: S - S_0 > 0$$

Then, the probability of error in detection, denoted by $PE^{(1)}$, can be written as

$$PE^{(1)} = 2Q\left(\frac{Cs}{\sqrt{2\left[(\sigma_n^2 + \eta^2 \sigma_r^2) + \frac{\sigma_\xi^2}{\Delta\rho^2}\right]}}\right) \quad (5)$$

Where $Q$ is the $Q-$ function defined as $Q(x) = \frac{1}{\sqrt{2\pi}} \int_x^\infty \exp\left(-\frac{u^2}{2}\right) du$ and $\eta = \left(\frac{\rho_b}{\rho_u} - 1\right)^{-1}$

The measurement noise in the system can be significantly reduced by conducting multiple experiments and averaging the results. For this case, the probability of error $\left(PE^{(M)}\right)$ with averaging over $M$ independent experiments becomes

$$PE^{(M)} = 2Q\left(\frac{Cs\sqrt{M}}{\sqrt{2\left[M(\sigma_n^2 + \eta^2 \sigma_r^2) + \frac{\sigma_\xi^2}{\Delta\rho^2}\right]}}\right) \quad (6)$$

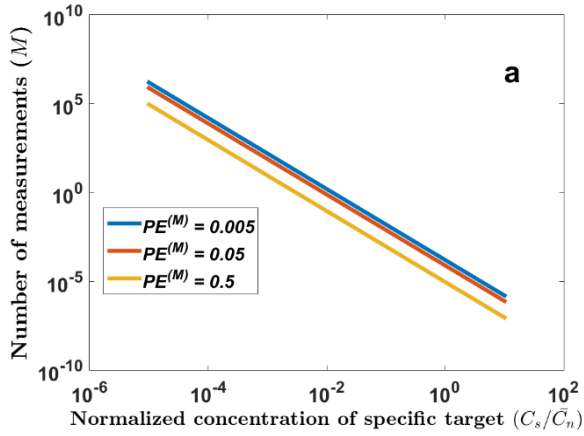
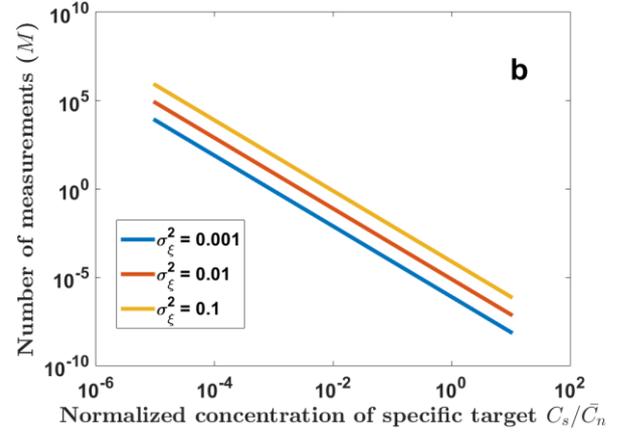
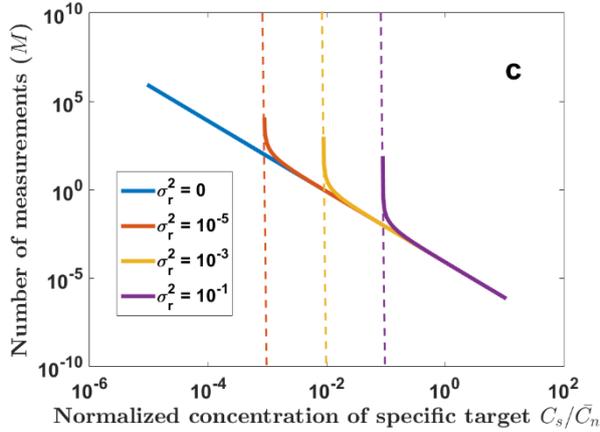
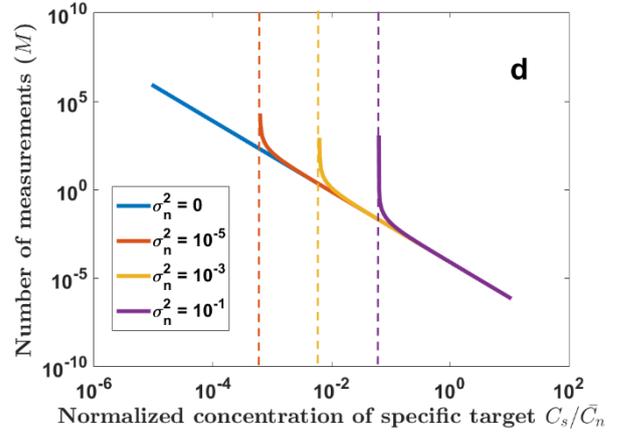

**Figure 2:** Effects of parameters on the number of measurements (M). (a) Effect of probability of error. Larger number of measurements is needed to achieve a lower probability of error $(PE^{(M)})$, [Parameters used: $\sigma_r^2 = \sigma_n^2 = 0, \Delta\rho = 10, \eta = 2, \sigma_\xi^2 = 0.1$] (b). Effect of measurement noise. Larger number of measurements are required to average out larger measurement noise. [Parameters used: $\sigma_r^2 = \sigma_n^2 = 0, \Delta\rho = 10, \eta = 2, PE^{(M)} = 0.05$] (c) and (d) Effect of fluctuations in the concentrations of receptor molecules and non-specific ligand molecules. The fluctuations in the concentrations creates a lower bound on the detectable specific molecule concentration which is equal to $(C_s/\overline{C_n})_{min}$ as defined in eq (8). [Parameters used: $\Delta\rho = 10, \eta = 2, \sigma_\xi^2 = 0.1, PE^{(M)} = 0.05$; For (c) $\sigma_n^2 = 0$ and for (d) $\sigma_r^2 = 0$]

**Results and Discussion:**

In a hypothesis testing problem, the degree of accuracy is defined by the user and hence, $PE$ is a user defined parameter and sets the extent of measurement fidelity where a low probability of error corresponds to high-fidelity measurement. We would like to explore how the detection limit scales with the choice of $PE$ as well as other experimental parameters. Concentration parameters, namely, $\overline{C_n}, \overline{C_r}$ can respectively be controlled by dilution of the serum and the amount of receptor molecules attached on the surface. The parameters $\rho_u$ and $\rho_b$ depend on the read-out technique. For example, in the case of a labelled detection technique based on fluorescently labeling the bound complexes, $\rho_u$ which represents the unlabeled (consequently non-fluorescent) population, would essentially be close to zero. On the other hand an unlabeled detection technique, for instance, relying on measurement of refractive index (RI) would result in comparable values of $\rho_u$ and $\rho_b$ due to the small magnitude of change (typically $10^{-5}$ - $10^{-6}$ RI units) expected due to binding [16]. We first ask the question how many independent measurements does one need to achieve a given probability of error $(PE)$ and how does this number scale with other experimental parameters such as uncertainty in receptor concentration? To answer this question, we modify equation (6) to a form

$$M = \frac{2(q^M)^2 \sigma_\xi^2}{\Delta\rho^2 (C_s^2 - 2(q^M)^2(\sigma_n^2 + \eta\sigma_r^2))} \qquad (7)$$

where $q^M = Q^{-1}(PE^{(M)}/2)$

Generally, one can reduce uncorrelated measurement noise by averaging over several independent measurements as shown in Figure 2(a) which shows that for any specified $PE^{(M)}$, arbitrarily low limits of detection of the target biomarker can be achieved by increasing the number of independent experiments. This finding supports the use of high-throughput sensing systems to achieve superior performance. Similarly, larger the magnitude of measurement noise $(\sigma_\xi)$, larger the number of measurements needed to achieve the specified error probability as shown in Fig 2(b). However, when there is uncertainty in the value of parameters $C_r$ and $C_n$, the number of independent measurements required to achieve a specified error probability diverges beyond a threshold value of detection limit (Fig. 2(c) and 2(d)). This implies that the detection limit has a practical lower bound and no amount of averaging over measurements can improve the limit of detection below this lower bound. The value of this lower bound depends on the level of uncertainty in the parameter values. Thus, our model predicts that arbitrarily low limits of detection of target biomarkers is not possible unless all the other parameter values are known with infinite precision. The absolute lower bound for the detection limit ($LoD$) for target biomarker concentration can be calculated by imposing the condition $M \to \infty$ in Eq. (7) to yield,

$$LoD = \left(\frac{C_s}{\overline{C_n}}\right)_{min} = \sqrt{2} q^{\infty} \sqrt{\left(\left(\frac{\eta \overline{C_r}}{\overline{C_n}}\right)^2 \delta_r^2 + \delta_n^2\right)} \quad (8)$$

The existence of a lower bound in detection limit resembles a similar result obtained in the context of communication systems [17]. In that work, the authors studied the effect of uncertainty in the assumed noise model on the performance of a radiometer used to detect the presence of a signal. They showed that if there is no uncertainty and the noise variance is completely known, then the detection error can be minimized by averaging the results of a large number of systems. Ideally, with the number of systems $(N) \to \infty$, one can detect the presence of the signal with infinite precision. Interestingly, if there is uncertainty in the modelled parameters, then there is a lower bound on the minimum detectable $SNR$, which they referred to as the "SNR Wall". Similar to this case, in our system, if there is no variation in the concentrations of non-specific and receptor molecules ($\sigma_r = \sigma_n = 0$) between independent measurements, any arbitrarily low concentration $C_s$ of the target biomarker can be detected using $M \sim 1/C_s^2$ independent measurements. However, similar to [17], when there are variations in the concentrations of non-specific and receptor molecules leading to uncertainty in the value of these parameters, there exists a "wall" which prevents achieving detection limits below a lower bound.

Another interesting aspect of equation (7) is the dependence of $M$ on the signal due to the unbound receptors ($\rho_u$). This aspect makes the system considered here different from the previous report on SNR wall in communication systems [17]. If the concentration of the receptors is known to infinite precision, i.e. $\sigma_r = 0$, then the number of measurements ($M$) depends only on the contrast between the signals due to bound receptors and unbound receptors ($\Delta\rho$) and not on $\rho_u$ explicitly. However, if $\sigma_r > 0$, then the number of measurements also depends on the ratio of the signals due to the bound and unbound receptors and therefore, there is an explicit dependency on the signal due to the unbound receptors $\rho_u$. This implies a significant difference between signal read-out techniques based on their response to the presence of unbound receptors. For instance, there are a class of sensing techniques which use secondary labels (e.g. a fluorophore) which bind on top of the bound receptors producing a fluorescent signal. In such cases, we can take $\rho_u$ to be 0. As shown in Fig. 3c, non-zero response to unbound receptors adds to the signal background and consequently for a larger $\rho_u$, a higher signal contrast ($\Delta\rho$) is required to achieve a given $LoD$. The parameter η characterizes the system's explicit dependence on the unbound receptors and hence controls the $LoD$. For a fixed value of $\Delta\rho$, a read-out technique with lower $\eta$ can achieve lower $LoD$ [Fig 3(a)]. On the other hand, for a given read-out technique (constant $\eta$), larger the signal contrast ($\Delta\rho$), lower the number of measurements required to detect a given concentration of specific target molecules [Fig 3(b)].

To validate results of our model system, we compared the $LoDs$ reported in the literature by dividing them into two categories namely labeled and label-free detection techniques [Fig 3d]. For each category, we calculated the cumulative distribution function ($CDF$) of the available data. To calculate the approximate $CDF$ from a set of observed data, we rank the observations in ascending order and use $CDF(x) = \frac{R(x)-1}{N-1}$, where R(x) is the rank (position) of x in the sorted list of observations and N is the total number of observations in the set. We then plot $CDF(x)$ against $x$ for each observed value in the set to obtain the $CDF$ curves shown in Fig. 3 (d) A metric referred as $LoD_{50}$ was then defined based on the $CDF$. The $LoD_{50}$ is the $LoD$ value for which the $CDF$ function reaches the mid-point. The $LoD_{50}$ value therefore represents a kind of weighted average of the $LoDs$ reported for each category. There were 54 data points for labeled detection and 53 data points for label-free detection making the comparison reasonable. Further, this comparison was done across proteins ranging from 10 kDa to 600 kDa with assays varying in the antibodies/receptors used, surface functionalization protocols and signal detection methods spanning mechanical, electrical, electrochemical and optical domains involving techniques as wide as Surface Plasmon Resonance, e.g. [18], micro-cantilevers, e.g. [19], fluorescence-immunoassays, e.g. [20], ELISA e.g. [21], optical interferometry, e.g. [22], and Silicon nanowires, e.g. [23]. Therefore, the data collected was a comprehensive representation of serum based labeled and label-free detection approaches. In order to extract the $LoD_{50}$ value, the data points were fitted with a smooth curve. As seen in Fig. 3 (d), the $LoD_{50}$ value of labeled detection was about 0.1 pM (pico-Molar) while that of label-free detection was around 10 pM indicating a 3 orders of magnitude gap in $LoDs$ in favor of labeled detection methods.

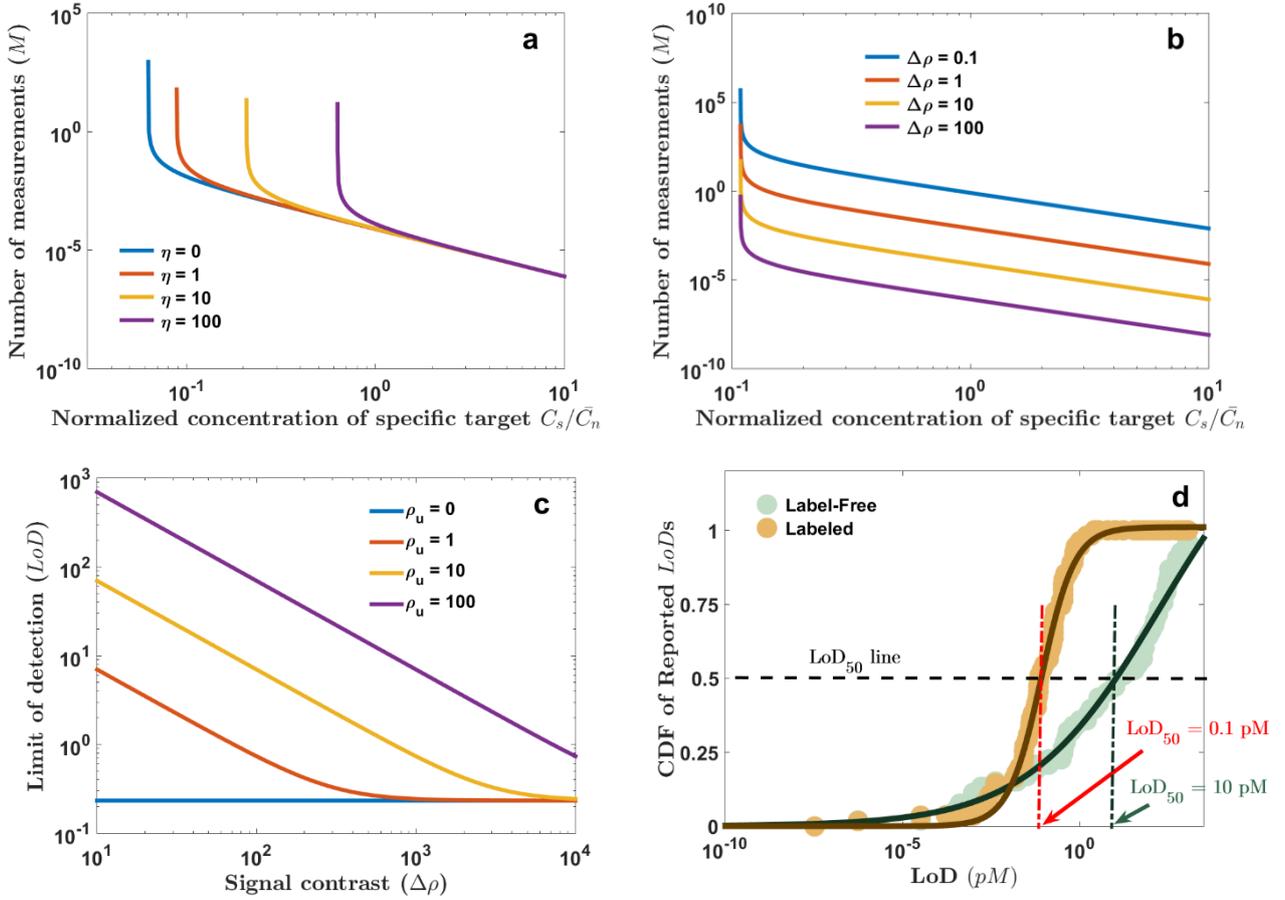

**Figure 3** Effect of signals due to bound and unbound receptors a. Effect of $\eta$ on the number of measurements. $\eta$ characterizes the signal due to unbound receptors and hence controls the $LoD$ For a given $\Delta\rho$, changing $\eta$ will shift the limit of detection. [Parameters used: $\Delta\rho = 10, \sigma_\xi^2 = 0.1, PE^{(M)} = 0.05, \sigma_r^2 = \sigma_n^2 = 0.1$] (b) Effect of signal contrast ($\Delta\rho$) on the number of measurements. With larger signal contrast, a smaller number of measurements are required to detect a given concentration. However, for a given $\eta$, increasing the signal contrast can't decrease the $LoD$. [Parameters used: $\eta = 2, \sigma_\xi^2 = 0. PE^{(M)} = 0.05, \sigma_r^2 = \sigma_n^2 = 0.1$] (c) Effect of signal contrast on the $LoD$ for different values of $\rho_u$. If $\rho_u = 0$, the $LoD$ can be achieved with any value of $\Delta\rho$ using appropriate number of measurements ($M$). Therefore, $\Delta\rho$ has no effect on the $LoD$ for $\rho_u = 0$. However, for $\rho_u > 0$, $LoD$ explicitly depends on $\Delta\rho$. [Parameters used: $PE = 0.025, \delta_r = 0.3, \delta_n = 0.1, \overline{C_r} = 10^{-3}, \overline{C_n} = 10^{-5}$ (d) $CDF$ of labeled and label-free $LoD$ data obtained from the literature. The $LoD_{50}$ which represents the weighted average shows that labeled detection techniques perform better than label-free techniques. The $CDF$ of labeled detection technique has $LoD_{50}$ two orders of magnitude lower than the $CDF$ of the label-free detection techniques]

We can now calculate the detection limit of an optimal detection system. As described earlier, the read-out technique of such a system must have $\rho_u = 0$. For this condition, the optimal limit of detection will become

$$(LoD)_{opt} = \sqrt{2} q^\infty \delta_n \qquad (9)$$

The optimal limit of detection only depends on the user defined probability of error ($PE^\infty$) and the uncertainty in the concentration of the non-specific target molecules in the serum. Figure 4 shows the effect of $PE^\infty$ on $LoD_{opt}$ for different values of $\delta_n$. We note that $PE^\infty$ does not affect the detection limit except at very large error probability i.e. close to 1. Typically, the desired probability of error is within the range 0.005 to 0.01, where there is practically no effect of $PE^\infty$ on $LoD_{opt}$. The parameter $\delta_n$ is an attribute of the serum and can't be controlled artificially. Therefore, under optimal conditions, the uncertainty in the concentration of non-specific target molecules present in the serum define the lower bound for detection of biomarkers.

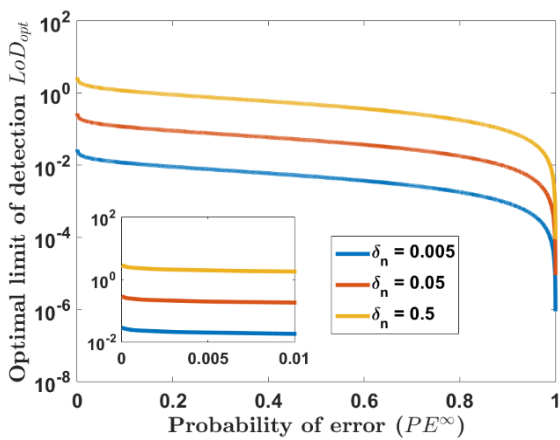

**Figure 4:** Effect of $PE^\infty$ on the optimal limit of detection $(LoD)_{opt}$ for different values of $\delta_n$. Practically $PE^\infty$ has no effect on the $LoD_{opt}$ except for very large error range. Therefore, $LoD_{opt}$ only depends on the uncontrollable parameter $\delta_n$.

**Conclusion:**

In a real-life environment, there are several sources of noise that hinder the performance of a sensing system. In case of clinical biosensors, there are typically three major sources of noise i.e. noise due to the measurement system and fluctuations in the concentrations of the receptors and the non-specific spurious ligand molecules in the serum. The noise due to the read-out process can be eliminated by performing large number of measurements and averaging the results. However, the noise due to the fluctuations in the concentrations can't be eliminated by ensemble averaging. Our analysis shows that the effects of the fluctuations in the receptor concentrations can be significantly reduced by using a measurement technique that has very low signal due to the unbound receptors. Therefore, one can achieve a lower $LoD$ by using read-out techniques that suppresses the signal due to the unbound receptors. This result justifies the experimental evidence that, fluorescence-based measurement techniques such as ELISA perform better than others. Our analysis will be useful for the systematic design of novel clinical biosensing techniques, specifically accounting for the measurement fidelity.